\documentclass[aps,prb,twocolumn,showpacs,superscriptaddress,groupedaddress,floatfix]{revtex4-1}  
\usepackage{graphicx} 
\usepackage{dcolumn}   
\usepackage{bm}        
\usepackage{amssymb}  

\everymath{\displaystyle}

\begin{document}
\newcommand{\kp}{{\bf k$\cdot$p}\ }
\newcommand{\Pp}{{\bf P$\cdot$p}\ }

\preprint{APS/123-QED}
\title{Cyclotron and combined phonon-assisted resonances in double-well heterostructure In$_{0.65}$Ga$_{0.35}$As/In$_{0.52}$Al$_{0.48}$As at megagauss magnetic fields }
\author{M. Zybert, M. Marchewka, and E. M. Sheregii}
 \affiliation{Centre for Microelectronics and Nanotechnology, University of Rzeszow, Rejtana 16a, 35-959 Rzeszow, Poland}
\author{P. Pfeffer, and W. Zawadzki}
 \affiliation{Institute of Physics, Polish Academy of Sciences,\\
Al.Lotnikow 32/46, 02-668 Warsaw, Poland}
\author{D. G. Rickel, J. B. Betts, F. F. Balakirev, M. Gordon, and C. H. Mielke}
 \affiliation{National High Magnetic Field Laboratory, Los Alamos, New Mexico 87545, USA}

\date{\today}
\begin{abstract}
Experiments on resonances of conduction electrons in InGaAs/InAlAs double quantum wells at megagauss magnetic fields in the Faraday geometry are reported. We observe new cyclotron resonances assisted by emission of InAs-like and GaAs-like optic phonons and a combined (cyclotron-spin) resonance assisted by emission of InAs-like phonon. The observations are very well described for three laser frequencies with the use of an eight-band (three level) $\textbf{k}\cdot \textbf{p}$ model, taking into account position- and energy-dependent effective masses and spin g-factors. The observation of  combined resonances in the Faraday configuration indicates that the energy subbands in the investigated quantum wells are not only markedly nonparabolic but also slightly nonspherical. It is indicated that the new observations are possible due to the application of very high magnetic fields.
\end{abstract}

\maketitle

\section{INTRODUCTION}

Magnetooptical transitions have been for years a powerful tool  in the investigations of  the bulk and two-dimensional (2D) semiconductor structures. In particular, they can supply  precise informations on the band structure and relevant band parameters. Magnetooptics is governed by well defined selection rules that give additional details of investigated systems and their symmetry. Because of their sharp resonant character, magnetooptical excitations allow to identify different physical objects , like impurities, plasmons, phonons, etc. The use of very high magnetic fields in experiments has important merits: they produce considerably stronger transitions and penetrate into high band energies often revealing additional physical subtleties.

In the present work we investigate with the help of magnetooptical resonances at megagauss pulsed fields a 2D system of two quantum wells made of mixed crystals InGaAs and InAlAs. The experiments are carried in the Faraday configuration with the use of laser light sources. The two InGaAs wells in our structure have relatively small energy gaps, so their description requires  a multi-band {\kp} theory with the effective masses and spin $g$-factors depending on the electron energies and positions.  In addition to the standard cyclotron resonances for spin-up and spin-down electron states, we observe a weaker combined cyclotron-spin resonance as well as all corresponding phonon-assisted resonances. The combined resonance and the phonon-assisted combined resonance were observed in bulk InSb by McCombe et al \cite{1} and Zawadzki et al \cite{2}, respectively. On the other hand, the observation of phonon-assisted cyclotron resonances as well as of  the combined and phonon-assisted combined resonances in 2D systems seem to have no precedent. We believe that these new observations have been possible due to the application of megagauss magnetic fields.

\section{EXPERIMENT}

The investigated structure was prepared by means of a low-pressure metal organic vapor phase epitaxy \cite{3} on the (100) plane semi-insulating InP substrate. It consisted of two In$_{0.65}$Ga$_{0.35}$As wells and three In$_{0.52}$Al$_{0.48}$As barriers, each doped by Si $\delta$ - layers. The well and barrier thicknesses were 20 nm. The 2D electron mobility in the wells was around 2.5 $\times 10^{5}$ cm$^{2}$/Vs at high electron density 3.5 $\times 10^{12}$ cm$^{-2}$, as measured by the magneto-transport at temperatures 1.6 - 4.2 K \cite{4,5}. Our optical experiments were carried out with the use of CO$_{2}$ laser radiation at three wavelengths: $\lambda_{1} = 10.59 \mu m$, $\lambda_{2} = 10.15 \mu m$ and $\lambda_{3} = 9.69 \mu m$, having the power of about 80 mW for all frequencies. The pulsed magnetic fields at the megagauss range were produced at the National High Magnetic Field Laboratory, Pulsed Field Facility in Los Alamos. The experimental set-up was described in details in our previous work \cite{6}. Magnetic fields up to 150 T were generated in a single-turn coil discharging a capacity of 250 kJ and inductance of 17.5 nH during $6 \mu s$. The magnetic field \textbf{B} was parallel to the growth direction of the wells and a special sample holder ensured the Faraday geometry. In order to detect the radiation transmitted through the sample, a HgCdTe detector was placed within the single turn coil. The modern system of the data acquisition provided registration of magneto-transmission for both increasing and decreasing magnetic field with resolution $1 Mega-samples/1 \mu s$. The magnetic field was measured at the sample using a dB/dt detecting coil, with an estimated uncertainty not exceeding $\pm 3 \% $.

Due to the high electron density, the contribution of 2DEG to the  magneto-optical data in the mid-IR range was the dominant one. The magneto-transmission spectra obtained at $T$ = 6 K are presented in Fig. 1 for three wavelengths of the incident radiation.  The strong transmission peak at 83 T for the $\lambda_{1}$ wavelength, at 85 T for $\lambda_{2}$ and at 86 T for $\lambda_{3}$ are seen  on the presented experimental curves. A minimum of magneto-transmission at these resonances reaches about 35 \%, which is the value of the absorption coefficient (calculated taking into account the value of thickness $d$ = 40 nm (both channels) and the reflection coefficient in middle infrared region $R=0.22$ \cite{4}) of the order $\alpha=10^{2}$ cm$^{-1}$.
The strongest peak corresponds to the cyclotron resonance (CR), related to the transition from
the lowest level 0$^+$ to the level 1$^+$. The weaker but still quite strong
peak is the CR for the transition 0$^-$ $\rightarrow$ 1$^-$. Much weaker transitions,
revealed by suppressing the noise with the use of Fourier transform
methods, are indicated by arrows in the insets for the appropriate light
wavelengths. Their field values are reported in figure 2.

\begin{figure}[hbp]
	\centering
		\includegraphics[width=0.45\textwidth]{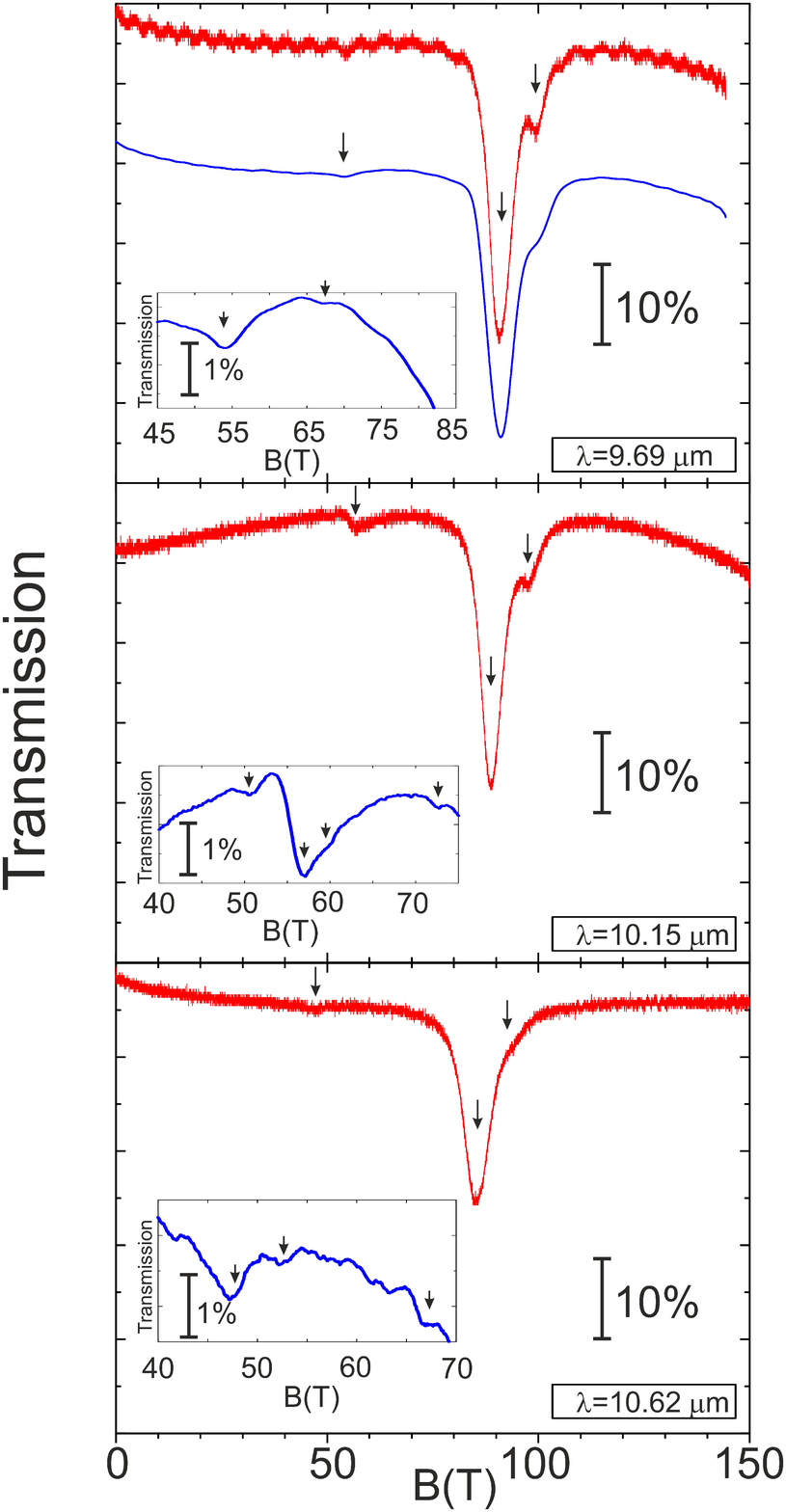}
\caption{Magneto-transmission spectra of 3183 DQW obtained at megagauss magnetic fields at 6 K for  three wavelengths of laser radiation: red curves - as recorded, blue ones - after filtering noise. Inserts show the parts of experimental curves in extended scale of transmission for narrow regions of magnetic field. The noise of data has been reduced using Fourier-transform procedures.}
\label{fig:fig1}
\end{figure}

\section{THEORY}

Since InGaAs alloys forming the wells in our system belong to medium-gap  materials, one should use for their description a multiband formalism of the {\kp} theory. Specifically, we use a  three-level  model (3LM) of  the {\kp} theory. We deal with wells and barriers along the growth direction $z$, so in our formalism the energy gaps and other band parameters are functions of $z$. The model takes into account eight bands (including spin) arising from  $\Gamma^v_7$,  $\Gamma^v_8$ (double degenerate) and  $\Gamma^c_6$ levels at the center of the Brillouin zone and treats the distant (upper and lower) levels as a perturbation. The resulting bands are spherical but nonparabolic (see however below). For the presence of a magnetic field the {\kp} theory becomes the {\Pp} theory and in 3LM (eight bands including spin) it takes the form \cite{7}

$$
\sum_l\left[\left( \frac{P^2}{2m_0}+E_{l}+V(z)-{\cal E}\right)\delta_{l'l} \right.
$$
\begin{equation}
\left.+\frac{1}{m_0}{\bf p}_{l'l}\cdot\bf{P} +\mu_{\rm B} {\bf B} \cdot{\bm \sigma}_{l'l} \right]f_l =0 \;\;,
\end{equation}
where ${\cal E}$ is the energy, $E_{l}$ are band edge energies at the center of the Brillouin zone, ${\bf P} = {\bf p}+e{\bf A}$ is kinetic momentum, ${\bf A}$ is vector
potential of magnetic field $\textbf{B}$, and ${\bm \sigma}_{l'l} = (1/\Omega)\langle u_{l'}|{\bm \sigma}|u_l\rangle$. Here ${\bm
\sigma}$ are the Pauli spin matrices, $\Omega$ is the volume of the unit cell, $u_l$ are periodic amplitudes of
the Luttinger-Kohn functions, $\mu_B = e\hbar/2m_0$ is the Bohr magneton, ${\bf p}_{l'l}$ are the interband
matrix elements of momentum. The sum runs over
all bands $l = 1, 2,...,8$ included in the model and $l'=1, 2,...,8$ runs over the same bands.
Within 3LM there exists the interband matrix element of momentum $P_0$ and that of the spin-orbit interaction $\Delta_0$ (see Ref. 8).
For two rectangular wells the potential $V(z)$ describes conduction band offsets and barriers. The valence offsets are automatically determined by corresponding energy gaps.

Equation (1) represents an 8$\times$8 system of equations for eight envelope functions $f_l(\textbf{r})$. Since we are interested in the eigenenergies
and eigenfunctions for the conduction band, we express the valence functions $f_3,...f_8$ by the conduction functions $f_1$ and $f_2$ for the spin-up and
spin-down states, respectively. The magnetic field $\textbf{B} || \textbf{z}$ is described by the asymmetric gauge $\textbf{A} = [-By, 0, 0]$. For above $\textbf{A}$, the envelope functions can be separated into the form $f_l = exp(ik_xx)\Phi_n[(y-y_0)/L]\chi_l(z)$, where $\Phi_n$ are the harmonic oscillator functions,
$y_0 = k_xL^2$ in which $L = (\hbar/eB)^{1/2}$ is the magnetic radius. After substituting the equations for l'=3, Ŭ 8 to the equations for l'=1, 2, the effective Hamiltonian for $f_1$
and $f_2$ functions becomes
\begin{equation}
\hat{H}=\left[ \begin{array}{cc}
\hat{A}^+&0\\
0&\hat{A}^-
\end {array}
 \right]\;\;,
\end{equation}
where
$$
\hat{A}^{\pm}=V(z)-\frac{\hbar^2}{2}\frac{\partial}{\partial z} \frac{1}{m^*( {\cal E},
z)}\frac{\partial}{\partial z}
$$
\begin{equation}
+\frac{\hbar\omega_c^0(n+1/2)}{2 m^*( {\cal E}, z)/m_0} \pm\frac{\mu_BB}{2}g^*( {\cal E}, z)
\;\;,
\end{equation}
in which $\omega_c^0=eB/m_0$ and $n$=0, 1, 2, Šis the Landau level number. The effective mass $m^*$ and the spin $g$-factor are
\begin{equation}
\frac{m_0}{m^*({\cal E}, z)}=1+C-\frac{1}{3}E_{P_0}\left(\frac{2}{\tilde E_0}+ \frac{1}{\tilde G_0}\right) \;\;,
\end{equation}
\begin{equation}
 g^*({\cal E}, z)=2+2C'+\frac{2}{3}E_{P_0}\left(\frac{1}{\tilde E_0}- \frac{1}{\tilde G_0}\right)\;\;,
\end{equation}
where ${\tilde E_0} = E_0 - {\cal{E}}+V(z)$ and ${\tilde G_0} = G_0 - {\cal{E}}+V(z)$. Here $G_0 = E_0 + \Delta_0$, $E_{P_0}=P^2_02m_0/\hbar^2$ and $C$ and $C'$ are far-band contributions. The effective masses and the $g$ values depend on the band structure and consequently are different for wells and barriers. They also depend on the energy due to bands' nonparabolicity. Values of band-edge parameters are given in Table 1 and discussed in Appendix.
For spherical energy bands, the transverse motion in $x-y$ plane is separated from that along the growth direction $z$. In consequence, this motion can be quantized into the Landau levels; this has been used in Eq.(3).

\begin{figure}[htbp]
	\centering
	\includegraphics[width=0.55\textwidth]{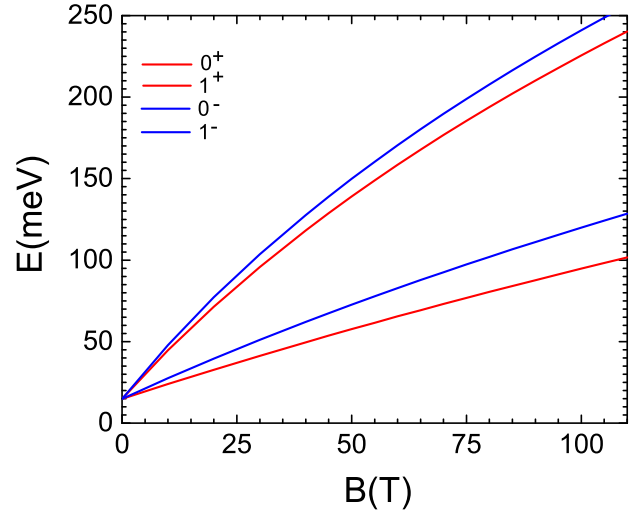}
	\caption{Calculated conduction Landau and spin levels in In$_{0.65}$Ga$_{0.35}$As/In$_{0.52}$Al$_{0.48}$As heterostructure involved in the experiments versus magnetic field. At $B = 0$ the levels converge to the lowest electric subband $e0 = 14.8$ meV.}
	\label{fig:fig2}
\end{figure}

Next the eigenenergy equations $\hat{A}^+ \chi^+={\cal{E}}^+_n \chi^+$ and $\hat{A}^- \chi^-={\cal{E}}^-_n \chi^-$ are solved separately along the $z$ direction for the two spin states using the boundary conditions
\begin{equation}
 \chi^{\pm}|_+ = \chi^{\pm}|_-,\;\;\;
 \left[\frac{1}{m^*}\frac{\partial \chi^{\pm}}{\partial z}\right]_+ = \left[\frac{1}{m^*}\frac{\partial \chi^{\pm}}{\partial z}\right]_-
\end{equation}
 at each interface. In addition one deals with the boundary conditions at $z = \pm\infty$, where the wave function must vanish.
  It is well known that, in a two-well system, the lowest electric subbands correspond to the symmetric and antisymmetric wave functions. The calculations indicate that, in our case, due to the wide potential barrier the energies corresponding to both wave functions are nearly the same and equal to $e0$ = 14.8 meV at $B$ = 0.
  
  In Fig. 2 we show the calculated energies  of the four lowest Landau states of our system as functions of  the external magnetic field. All four lines converge at $B$=0 at the electric subband energy $e0$. It is seen that the magnetic energies are sublinear functions of $B$, which is a manifestation of bands' nonparobolicity resulting from a rather small value of the energy gap in In$_{0.65}$Ga$_{0.35}$As, see Table 1.

\begin{table}
\caption{Band parameters of alloys for different chemical compositions $x$, as used in the
calculations, $C$ and $C'$ are far-band contributions to the band-edge values.
Interband matrix element of momentum $P_0$ is taken to be independent of $x$: $E_{P_0}$ = 27 eV. The offset is $V_B$ = 0.50 eV. Energies of optic phonons: $\hbar\omega_{L1}$ = 28.64 meV for InAs and $\hbar\omega_{L2}$ = 33.23 meV for GaAs.
}

\begin{ruledtabular}

\begin{tabular}{ccc}
&In$_{.65}$Ga$_{.35}$As&In$_{.52}$Al$_{.48}$As\\
\hline
 E$_0$(eV) & -0.723 & -1.457  \\
G$_0$(eV)&-1.078&-1.806\\
C&-13.84&-5.99\\
C'&-0.4&+0.61\\
m$^*_0$/m$_0$&0.049&0.081\\
g$^*_0$&-7.0&+0.83\\
\end{tabular}
\end{ruledtabular}
\end{table}

\section{RESULTS AND DISCUSSION}

Figure 3 shows our results for the experimental magneto-optical transitions compared to theoretical description. In the same figure we indicate initial and final states of the corresponding transitions. Thus, going from right to left, one deals with two CR $0^+\rightarrow1^+$ and $0^-\rightarrow1^-$ for the three laser wavelengths. It is seen that, with our choice of parameters, we obtain a very good description of the data.

\begin{figure}[htp]
	\centering
		\includegraphics[width=0.55\textwidth]{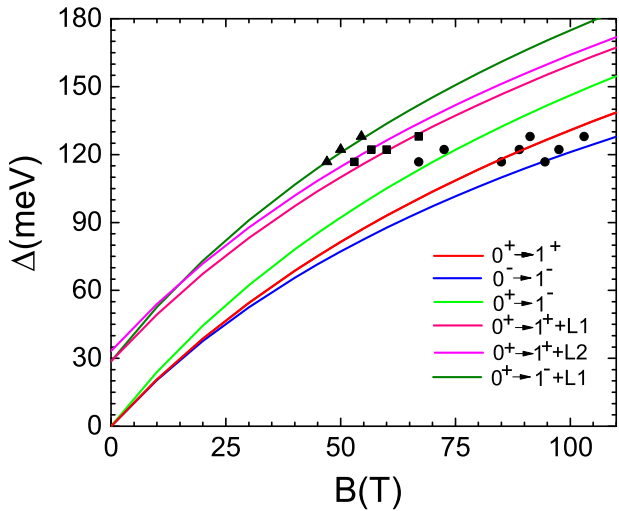}
\caption{\label{fig:epsart}{Experimental values of observed magneto-optical resonances for three frequencies of CO$_2$ laser versus magnetic field, as compared with the theoretical description. The Landau and spin levels involved in the transitions are indicated. The three highest theoretical lines correspond to two phonon-assisted
cyclotron resonances and a phonon-assisted combined resonance, respectively.}} 
	\label{fig:fig3}
\end{figure}

Two resonances at magnetic fields between 67 and 73 T, of a much weaker intensity, correspond, as far as the energies are concerned,  to the combined cyclotron-spin resonance (CBR): $0^+\rightarrow 1^-$, for which the Landau number changes by one and the spin is flipped. The resonance condition for this case is
$\hbar\omega = {\cal{E}}_1^- - {\cal{E}}_0^+$,
where $\hbar\omega$ is the photon energy. In this transition the change of the total angular momentum $\Delta j = \Delta l+\Delta s$ is zero because $\Delta l$ = +1, while $\Delta s$ = -1. Such a  transition should not be possible in a spherical band for the Faraday configuration since in this case photons carry the angular momentum +1 or -1 on the direction of their propagation. The CBR transition is possible in the Voigt configuration for the light polarization parallel to magnetic field. This selection rule may be obtained by a direct calculation of the corresponding matrix element \cite{1,9,10}. Thus, our three-level {\Pp} description, resulting in a spherical conduction band, does not account for this transition.
 Nevertheless, the CBR transition in our situation can be explained in several ways. The first is experimental: if the directions of light propagation and that of magnetic field are not perfectly parallel, there exists a nonvanishing component of light polarization on the the field direction which makes it possible to excite the observed transition. Further, the above angular momentum considerations are strictly valid only for a spherical conduction band. However, the conduction bands in III-V compounds are not truly spherical. Due to the {\kp} interaction with distant bands the $\Gamma^v_8$ valence and the  $\Gamma^c_6$  conduction bands are warped. This was shown explicitly for InSb and InAs  \cite{11,12} and for GaAs \cite{13}. In consequence, motion in the $x-y$ plane can not be completely separated from that along the growth direction $z$ and the CBR transition becomes possible.
 The nonsphericity of the conduction bands in III-V compounds exists also due the inversion asymmetry of the crystals which is reflected in the spin resonance \cite{14}. Last but not least, InGaAs/InAlAs heterojunctions on InP substrate are strained  \cite{15}. The presence of strain breaks the cubic symmetry of the crystal, giving rise to spin-orbit terms in the Hamiltonian which are linear in the electron wavevector and components of the strain. This leads to a strong enhancement of the spin flip transitions \cite{16,17}. In addition, the very high magnetic fields  strongly enhance CBR intensity since  the latter is proportional to $B^{3/2}$ \cite{11}. All in all, we are lead to the conclusion that a weak combined cyclotron-spin resonance is possible in our experiments (see also comments below).

 According to the calculated energies, four experimental points seen at  fields 50 - 67 T  (marked by squares) correspond to two cyclotron resonance transitions accompanied by emissions of InAs-like and GaAs-like optic phonon, respectively. The resonance condition in this
case is: $\hbar\omega = {\cal{E}}_1^{\pm} - {\cal{E}}_0^{\pm} + \hbar\omega_{L}$. Phonon-assisted spin-conserving magnetooptical transitions, first proposed by Bass and Levinson \cite{18}, were observed in bulk semiconductors on various occasions, see e.g. Ref. 19. However, it appears that the cyclotron phonon-assisted resonances have not been observed as yet on 2D systems. Since the intensity of CR transitions is proportional to B, the observability of the above resonances is greatly facilitated by the high fields used in the experiments.

Finally, according to the calculated energies, the three points around 50 T (marked by triangles) correspond to the phonon-assisted CBR. The resonance condition in this case is: $\hbar\omega = {\cal{E}}_1^- - {\cal{E}}_0^+ + \hbar\omega_{L1}$.
 As mentioned above, the  phonon-assisted CBR was observed in the photoconductivity on bulk InSb \cite{2}. Remarkably, the bulk data in Ref. 2 show a weak but nonvanishing resonance peak of the phonon-assisted CBR also for the light polarization transverse to \textbf{B}. This in addition strongly supports our considerations given above for the CBR not assisted by phonons.

The group of three experimental points seen in Fig. 3 at around 50 T (marked by triangles) can be, as far as the energies are concerned, explained alternatively as a CBR in which, instead of an optic phonon $\hbar\omega_{L1}$ emission, an electron goes simultaneously from the lowest electric subband to the first excited subband. The resonance energy in this case is also given by by the condition given above in which
 the phonon energy is replaced by the energy difference e1 - e0 between the two electric subbands. It so happens that the intersubband energy and the optic phonon energy are almost equal: e1 - e0 = 40.2 meV at $B$ = 0.

However, an electron transition between two electric subbands in the Faraday configuration is also problematic. The reason is that, as follows from elementary theoretical considerations, for a simple parabolic and spherical band the matrix elements of optical transitions between electric subbands vanish for the light polarization perpendicular to the growth direction $z$. As follows from a detailed analysis \cite{20}, such intersubband transitions become possible if one includes in the theory interband {\kp} mixing which occurs in narrow-gap semiconductors and is accompanied by band's nonparabolicity. The alloy In$_{0.65}$Ga$_{0.35}$As is a medium-gap semiconductor and, as follows from the sublinear $B$ dependence of the calculated energies shown in Fig. 2, we indeed deal with band's nonparabolicity. Thus optical intersubband transitions in the Faraday configuration should be weakly allowed in our case. Estimations in Ref. 20, indicate that intensities of intersubband transitions for the light polarization perpendicular to the growth direction are roughly ten times smaller than those for the parallel polarization. As it stands, we give preference to the phonon-assisted interpretation since it is consistent with our other observations, but this problem requires further studies.

Generally speaking, the observed phonon-assisted transitions are much
weaker than the "straight" ones without phonon assistance in agreement
with the theoretical predictions. In order to make the weak resonances
stronger one needs to employ laser lines of higher energies and observe
the resonances at still higher magnetic fields.

Interaction of optic phonons with 2D electrons in a magnetic field was studied previously in GaAs/GaAlAs heterostructures by means of magneto-polaron effects strongly affecting CR energies near the resonance condition $\hbar\omega_{c} = \hbar\omega_{L}$ \cite{21,22,23}.  Our study is also related to the electron-phonon interaction but the observed effects are different. Peeters et al. \cite{24} calculated the influence of electron-phonon interaction on the effective mass of 2D electrons. The mass used in our description is a bare effective mass because at megagauss fields  $\hbar\omega_{c} >> \hbar\omega_{L}$ and the polaron effects are negligible.

\section{SUMMARY}

In summary, we carried magnetooptical studies on double-well In$_{0.65}$Ga$_{0.35}$As/In$_{0.52}$Al$_{0.48}$As heterostructure for three laser frequencies at magagauss magnetic fields in the Faraday geometry. Two cyclotron resonances are observed which were successfully described with the use of multiband {\kp} theory. A weaker resonance is observed at lower magnetic fields (corresponding to higher energy transition) which we identify as a combined cyclotron-spin excitation. Possibility of such a transition in the Faraday geometry is related to a nonsphericity of the conduction band. Resonances observed at still lower fields are identified as cyclotron excitations assisted by emission of optic phonons. Finally, the highest energy excitation observed for all three frequencies is identified as a combined resonance assisted either by an optic phonon emission or by a transition between two electric subbands. The observation of weak magnetooptical resonances representing novel features of our work is possible owing to the very high magnetic fields employed in the experiments.

\begin{acknowledgments}
This work is supported by National Science Foundation - Cooperative Agreement No. DMR-1157490, the State of Florida, and the U.S. Department of Energy. We are grateful to NSF and LANL for the opportunity to perform the CR experiment.
\end{acknowledgments}

\appendix*
\section{}

Below we discuss parameters used in our calculations and compare them with the results of other authors. One should bear in mind that In$_{1-x}$Ga$_x$As wells are decisive for the electron energies, so we concentrate our attention on their features. Further, it is useful to remember that, for the above chemical  composition $x$ = 0.35, one is roughly "one third of the way" from InAs and "two thirds of the way" from GaAs. Below we quote, among others, the recommended  values from the review work of Vurgaftman et al \cite{25} and refer to them as [rcm].

We begin with the band-edge effective masses m$^*_0$/m$_0$. For GaAs there is 0.066 \cite{13} and 0.067 - 0.07 [rcm]. For InAs there is 0.026 [rcm]. The interpolated value  for our composition would be 0.040. There exist various measurements for the alloys of our interest  $x$ = 0.35 with the results: 0.033 \cite{26}, 0.037 \cite{27}, 0.038 \cite{28}, 0.046 \cite{29}. Thus the dispersion is quite large.  Our adjusted value is 0.049 (see Table 1). It can be seen from Eq. (4) that one can obtain a given value of m$^*_0$/m$_0$ by adjusting $C$ and $E_{P_0}$ at ${\cal E}$ = 0. These parameters are not equivalent since $C$ contributes only to the band-edge value of the mass, while $E_{P_0}$ determines also its energy dependence. The determined values of $E_{P_0}$ for InAs (in eV) are 21.5 - 22.2 and the rcm value is 21.5. For GaAs, the determined values (in eV) are 25.5 - 29 and rcm value is 28.8. Our adjusted value for x = 0.035 is 27 eV, which is somewhat too high. We can get an equally good fit to the data taking a lower value $E_{P_0}$ = 22 eV, but it would require the band-edge mass 0.0535 further away from the estimations of other authors.

The above mass values are related  to the band structure, whereas in principle one should take into account also the polaron corrections related to the interaction with optic phonons. At sufficiently high magnetic fields, when the cyclotron frequency becomes higher than the phonon frequency, the polaron corrections disappear. InAs and GaAs are weakly polar materials having the polar coupling constants $\alpha$ (InAs) = 0.0107 \cite{30} and $\alpha$ (GaAs) = 0.05 \cite{30}, so one can ignore the polaron corrections.

As to the band-edge spin $g$-values, one has for InAs g$^*_0$ = -14.7 \cite{31} and for GaAs g$^*_0$ = -0.44. For In$_{1-x}$Ga$_x$As with $x$ = 0.47 one finds g$^*_0$ = -3.95 \cite{32} and - 4.1 \cite{33}. Thus our adjusted value g$^*_0$ = -7.0 is quite reasonable.

All in all, taking into account uncertain characteristics of our sample (most notably the stress) and quite large dispersion of the mass values quoted above, we consider that the adjusted values of the effective mass and the spin  g$^*_0$ values given in Table 1 are reasonable. Since they have been adjusted to fit the high-field  data in the wide range of 50 - 105 T, they should be treated as a legitimate contribution to the characterization of In$_{0.65}$Ga$_{0.35}$As alloys.

\end{document}